\documentclass[12pt]{article}
\usepackage{amsmath,amssymb,bm,graphicx}
\usepackage{color} 
\usepackage{hyperref}

\newcommand{\pslash}{\not \! p}

\begin{document}

\vskip 0.5 truecm

\begin{center}
{\Large{\bf  Lorentz invariant CPT breaking\\in the Dirac equation\footnote{Talk given by one of the authors (KF) at the Memorial Meeting for  Abdus Salam's 90th Birthday, Nanyang Technological University, Singapore,  January 25-28, 2016 (to appear in the Proceedings).} 
}}
\end{center}
\vskip .5 truecm
\begin{center}
{\bf { Kazuo Fujikawa$^\dagger$ and Anca Tureanu$^*$}}
\end{center}

\begin{center}
\vspace*{0.4cm} 
{\it {$^\dagger$ Quantum Hadron Physics Laboratory, RIKEN Nishina Center,\\
Wako 351-0198, Japan\\
$^*$Department of Physics, University of
Helsinki, P.O.Box 64, \\FIN-00014 Helsinki,
Finland\\
}}
\end{center}


\begin{abstract}
If one modifies the Dirac equation in momentum space to 
$[\gamma^{\mu}p_{\mu}-m-\Delta m(\theta(p_{0})-\theta(-p_{0})) \theta(p_{\mu}^{2})]\psi(p)=0$, the symmetry of positive and negative energy eigenvalues is lifted by $m\pm \Delta m$ for a small $\Delta m$.  The mass degeneracy of the particle and antiparticle is thus lifted in a Lorentz invariant manner since the combinations $\theta(\pm p_{0})\theta(p_{\mu}^{2})$ with step functions are manifestly Lorentz invariant.  We explain an explicit  construction of this CPT breaking term  in coordinate space, which is Lorentz invariant but non-local at a distance scale of the Planck length.
The application of this Lorentz invariant CPT breaking mechanism to the possible mass splitting of the neutrino and antineutrino in the Standard Model is briefly discussed.

\end{abstract}


\section{Introduction}

It is timely to have a memorial meeting for the 90th birthday of the late Professor Abdus Salam when those who directly worked with him are still  active in physics. On the part of one of the present authors (KF), he first met Salam at  a small gathering in the suburb  of Oxford in 1972. This was a nice gathering, and M. Veltman talked on a work with G. 't Hooft and KF also talked on the so-called $R_{\xi}$ gauge written with B.W. Lee and A.I. Sanda.   Salam gave a talk on his famous unification paper. At that time, people were talking about the mass value of the W boson at least as heavy as 37.5 GeV; this value, which is not large by the present standard, 
was regarded to be absurdly large at that time. Note that this was before the discovery of the charmed quark. To defend this large mass needed in his model,  Salam was saying, ``if you insist on simplicity, Newton's theory of gravity which requires a single scalar potential is much simpler than Einstein's theory which requires 10 metric variables. But we prefer Einstein's theory because of  its beautiful symmetry.  The decisive factor is not superficial simplicity but  rather deep symmetry behind any theory''.  KF was very impressed by his verdict and still remembers it.
\\

The greatest contribution of  Salam is probably the proposal of the Standard Model, namely,  
the Weinberg-Salam-Glashow Model or Weinberg-Salam theory with Higgs mechanism~\cite{weinberg, salam, glashow},  among his numerous 
contributions to physics. 
The Standard Model is very successful so far, and people discuss 
the possible extensions of the model.
We would like to present a possible breaking scheme of CPT symmetry  in a Lorentz invariant non-local theory which produces a particle-antiparticle mass splitting,  and   as an application of this CPT breaking scheme, we briefly discuss the possible mass splitting between the neutrino and antineutrino in a minimal extension of 
the Standard Model. 
\\

We start with the {\bf CPT Theorem:} 
This theorem states that CPT symmetry is valid for any Lorentz invariant local theory defined by a  hermitian Lagrangian with normal spin-statistics relation.

The original references on the CPT theorem are~\cite{pauli, luders},\\
W. Pauli, in {\it Niels Bohr and the Development of Physics}, W. Pauli
(ed.), Pergamon Press, New York, 1955,\\
and \\
G. L\"{u}ders, {\it On the Equivalence of Invariance under
Time-Reversal and under Particle-Antiparticle Conjugation for
Relativistic Field Theories},
Det. Kong. Danske Videnskabernes
Selskab, Mat.-fys. Medd. {\bf 28} (5) (1954).

The title of this second reference shows clearly what we mean by the CPT theorem, namely, the equivalence of PT symmetry ({\em strong reflection})  with C symmetry. 
More generally, CPT symmetry implies the degeneracy of the masses of the particle and antiparticle. The CPT theorem, which is valid for any sensible local Lorentz invariant theory,  is very general. Nevertheless the possible breaking of CPT theorem
has been discussed by many people in the past. 
One may count
two representative necessary conditions to break the CPT theorem:\\
1. Non-local theory,\\
2. Lorentz symmetry breaking.\\

The Lorentz symmetry breaking scheme is more common in  literature (see, for example, \cite{piguet} and references therein),  but we are interested in the possible breaking of CPT in a Lorentz invariant non-local theory. To be specific, we mainly discuss the modified Dirac equation
\begin{eqnarray}
[\pslash-m-\Delta m(\theta(p_{0})-\theta(-p_{0})) \theta(p_{\mu}^{2})]\psi(p)=0
\end{eqnarray}
with the step function $\theta(x)=1$ for $x>0$ and $\theta(x)=0$ for $x<0$. The combinations $\theta(\pm p_{0})) \theta(p_{\mu}^{2})$ are manifestly Lorentz invariant since the signature of the time-component of any time-like vector has a Lorentz invariant meaning.  Obviously, this modified Dirac equation produces the mass splitting between the particle and antiparticle, $m\pm \Delta m$ for small $\Delta m$, and thus leads to the breaking of CPT symmetry. In passing, we emphasize that the breaking of CPT symmetry is a necessary condition but not sufficient for the particle-antiparticle mass splitting. For example, T breaking with C and P intact leads to CPT breaking but no mass splitting between the particle and antiparticle. 

We explain how to reproduce the equation (1) at energy scale $E/M_{P}\ll 1$  as a  consequence of the Lorentz invariant  CPT breaking which is non-local at the Planck length $1/M_{P}$~\cite{FT}. As is clear in the statement of the CPT theorem, we have to sacrifice some basic assumptions in  current physics to realize CPT symmetry breaking.  We discuss the interesting aspects of the present non-local CPT symmetry breaking scheme as well as the remaining issues to be resolved.  In this connection, it may be appropriate to mention that the 
more common CPT symmetry breaking on the basis of Lorentz symmetry breaking also encounters severe difficulties once one attempts to realize the mass splitting between the particle and antiparticle~\cite{piguet}.

\section{Lorentz invariant CPT breaking with \\particle-antiparticle mass splitting}

We have studied the models of Lorentz invariant CPT breaking  
which give rise to the particle and antiparticle mass splitting in the Dirac equation in a series of papers~\cite{CFT1,CFT2,FT}. We would like to recapitulate the essence of these works.

We start with a free Dirac action
\begin{eqnarray}\label{(2.1)}
S=\int d^{4}x\bar{\psi}(x)[i\gamma^{\mu}\partial_{\mu}-m]\psi(x),
\end{eqnarray}
and consider to add a small Lorentz invariant non-local term to break CPT symmetry.  We examine the 
hermitian combination with a real $\mu$~\cite{CFT1},
\begin{eqnarray}\label{(2.1)}
\int d^{4}xd^{4}y[\theta(x^{0}-y^{0})-\theta(y^{0}-x^{0})]\delta((x-y)^{2}-l^{2})[i\mu\bar{\psi}(x)\psi(y)].
\end{eqnarray}
The transformation property of the operator part in this expression is given  using {\em spin-statistics theorem} by,
\begin{eqnarray}
&& {\rm C}:\ i\mu\bar{\psi}(x)\psi(y)\rightarrow i\mu\bar{\psi}(y)\psi(x),\nonumber\\
&& {\rm P}:\ i\mu\bar{\psi}(x^{0},\vec{x})\psi(y^{0},\vec{y})\rightarrow
i\mu\bar{\psi}(x^{0},-\vec{x})\psi(y^{0},-\vec{y}),\nonumber\\
&& {\rm T}:\ i\mu\bar{\psi}(x^{0},\vec{x})\psi(y^{0},\vec{y})\rightarrow
-i\mu\bar{\psi}(-x^{0},\vec{x})\psi(-y^{0},\vec{y}),
\end{eqnarray}
and the overall transformation property of the combination in (3) is confirmed to be
\begin{eqnarray}
C=-1,\  P=1,\   T=1.
\end{eqnarray} 
Namely, 
\begin{eqnarray}
 C=CP=CPT=-1,
 \end{eqnarray}
and thus all symmetries which may protect the equality of the masses of the particle and antiparticle are broken.

It is thus interesting to examine a Lorentz invariant and hermitian action~\cite{CFT1}
\begin{eqnarray}
S&=&\int d^{4}x\{\bar{\psi}(x)i\gamma^{\mu}\partial_{\mu}\psi(x)
 - m\bar{\psi}(x)\psi(x)\nonumber\\
 && -\int
d^{4}y[\theta(x^{0}-y^{0})-\theta(y^{0}-x^{0})]\delta((x-y)^{2}-l^{2})[i\mu\bar{\psi}(x)\psi(y)]\}.
\end{eqnarray}
The Dirac equation is replaced by
\begin{eqnarray}\label{(2.4)}
&&i\gamma^{\mu}\partial_{\mu}\psi(x)=m\psi(x)+i\mu\int
d^{4}y[\theta(x^{0}-y^{0})-\theta(y^{0}-x^{0})]\delta((x-y)^{2}-l^{2})\psi(y).
\end{eqnarray}
By inserting an ansatz for the possible solution
\begin{eqnarray}\label{(2.5)}
\psi(x)=e^{-ipx}U(p),
\end{eqnarray}
we have
\begin{eqnarray}\label{(2.6)}
\pslash U(p)&=&mU(p)\nonumber\\
&+&i\mu\int d^{4}y[\theta(x^{0}-y^{0})-\theta(y^{0}-x^{0})]\delta((x-y)^{2}-l^{2})e^{-ip(y-x)}U(p)\nonumber\\
&=&mU(p)
+i\mu[f_{+}(p)-f_{-}(p)]U(p),
\end{eqnarray}
where 
\begin{eqnarray}\label{(1.3)}
&&f_{\pm}(p)\equiv \int d^{4}z_{1}
e^{\pm ipz_{1}}\theta(z_{1}^{0})\delta((z_{1})^{2}-l^{2}),
\end{eqnarray}
is a Lorentz invariant form factor.

For the {\em space-like} $p$,  we go to the frame where $p_{0}=0$, and we have  $f_{+}(p)=f_{-}(p)$ and no mass splitting, and thus 
no tachyons.

For the {\em time-like} $p$, we go to the frame where $\vec{p}=0$,
and the eigenvalue equation 
\begin{eqnarray}\label{(2.9)}
p_{0}&=&\gamma_{0}\{m + i\mu [f_{+}(p_{0})-f_{-}(p_{0})]\},
\end{eqnarray}
is written as
\begin{eqnarray}\label{(2.10)}
p_{0}&=&\gamma_{0}\left[m - 4\pi \mu\int_{0}^{\infty}dz\frac{z^{2}\sin [
p_{0}\sqrt{z^{2}+l^{2}}]}{\sqrt{z^{2}+l^{2}}}\right],
\end{eqnarray}
where we used the explicit form in (11).

This eigenvalue equation under $p_{0}\rightarrow -p_{0}$ becomes
\begin{eqnarray}\label{(2.11)}
-p_{0}
&=&\gamma_{0}\left[m + 4\pi\mu\int_{0}^{\infty}dz\frac{z^{2}\sin [
p_{0}\sqrt{z^{2}+l^{2}}]}{\sqrt{z^{2}+l^{2}}}\right].
\end{eqnarray}
By sandwiching this equation by $\gamma_{5}$,  we have
\begin{eqnarray}\label{(2.13)}
p_{0}
&=&\gamma_{0}\left[m + 4\pi\mu\int_{0}^{\infty}dz\frac{z^{2}\sin [
p_{0}\sqrt{z^{2}+l^{2}}]}{\sqrt{z^{2}+l^{2}}}\right],
\end{eqnarray}
which is not identical to the original equation in (13).

In other words, if $p_{0}$ is the solution of the original equation,
$-p_{0}$ cannot be the solution of the original equation for $\mu\neq 0$.
The last term in the Lagrangian (7) with
C=CP=CPT=$-1$ thus splits the particle and antiparticle masses.
As a crude estimate of the mass splitting, one may assume that the term with  $\mu$ gives a much smaller contribution than 
$m$ and solve these equations iteratively. If the particle mass is chosen at
\begin{eqnarray}\label{(2.14)}
p_{0}\simeq m - 4\pi \mu\int_{0}^{\infty}dz\frac{z^{2}\sin [ m
\sqrt{z^{2}+l^{2}}]}{\sqrt{z^{2}+l^{2}}},
\end{eqnarray}
then the antiparticle mass is estimated at
\begin{eqnarray}\label{(2.15)}
p_{0}\simeq m + 4\pi \mu\int_{0}^{\infty}dz\frac{z^{2}\sin [ m
\sqrt{z^{2}+l^{2}}]}{\sqrt{z^{2}+l^{2}}}.
\end{eqnarray}

We here mention that the factor $f_{\pm}(p)$ in (11) is mathematically related to
the two-point Wightman function,
\begin{eqnarray}\label{(13)}
\langle 0|\phi(x)\phi(y)|0\rangle=\int d^{4}p
e^{i(x-y)p}\theta(p^{0})\delta(p^{2}-m^{2}),
\end{eqnarray}
if one replaces the coordinate and momentum, $x^{\mu}\leftrightarrow p^{\mu}$.
This knowledge is useful to analyze our problem.  For example, the Wightman function is quadratically divergent at short distances, $(x-y)^{\mu}\rightarrow 0$, which implies that our form factor is quadratically divergent in the infrared, $p^{\mu}\rightarrow 0$. To avoid this infrared divergence, we replace the non-local factor in (3) to~\cite{FT}
\begin{eqnarray}
\delta\left((x-y)^{2}-l^{2}\right)\Rightarrow \Delta_{l}(x-y)\equiv \delta\left((x-y)^{2}-l^{2}\right)-\delta\left((x-y)^{2}-{l^{\prime}}^{2}\right)
\end{eqnarray}
with ${l^{\prime}}^{2}\rightarrow 0$ in practical applications.

With this replacement, 
for  time-like $p^{2}>0$, one may obtain in the frame $\vec{p}=0$,
\begin{eqnarray}\label{(14)}
p_{0}&=&\gamma^{0}[m+f(p_{0})],
\end{eqnarray}
with
\begin{eqnarray}\label{(15)}
f(p_{0})
&\equiv&i[f_{+}(p_{0})-f_{-}(p_{0})]\nonumber\\
&=&-4\mu\pi\int_{0}^{\infty}dz\Big\{\frac{z^{2}\sin [p_{0}\sqrt{z^{2}+l^{2}}]}{\sqrt{z^{2}+l^{2}}}
-\frac{z^{2}\sin [p_{0}\sqrt{z^{2}}]}{\sqrt{z^{2}}}\Big\}.
\end{eqnarray}
For space-like $p^{2}<0$, one can confirm that the $CPT$ violating term vanishes, $f(p)=0$, by choosing $p_{\mu}=(0, \vec{p})$.\\

\noindent {\bf Evaluation of mass splitting}\\

The CPT breaking factor in (21), which is now written as $f(p)$ for simplicity, is rewritten as~\cite{FT}
\begin{eqnarray}\label{(20)}
f(p)
&=&4\pi\mu l^{2}[\theta(p_{0})-\theta(-p_{0})]\theta(p^{2})\nonumber\\
&&\times\Big\{\int_{1}^{\infty}du \frac{1}{2u(\sqrt{u^{2}-1}+u)^{2}}\sin (|p_{0}|l u)\nonumber\\
&&-\frac{1}{2}\int_{0}^{1}du \frac{\sin (|p_{0}|l u)}{u} +\int_{0}^{1}du u
\sin (|p_{0}|lu) +\frac{1}{2}\int_{0}^{\infty}du \frac{\sin (u)}{u}\Big\},
\end{eqnarray}
namely, the $CPT$ violating term is characterized by the quantity  
\begin{eqnarray}\label{(21)}
\mu l^{2},
\end{eqnarray}
which has the dimensions of mass.
If one chooses the non-locality length $l$ at the Planck length,
we have $|p_{0}|l\ll 1$ for the energy scale at laboratory, and the formula (22) gives a manifestly Lorentz invariant
\begin{eqnarray}\label{(22)}
f(p)
&\simeq&\pi^{2}\mu l^{2}[\theta(p_{0})-\theta(-p_{0})]\theta(p^{2}).
\end{eqnarray}
Note that no term linear in $|p_{0}|l$ arises.
Thus the particle-antiparticle mass splitting is given by 
\begin{eqnarray}\label{(23)}
2\Delta m &\simeq&2\pi^{2}\mu l^{2}.
\end{eqnarray}
Our $CPT$ violating term $f(p_{0})$ is odd in $p_{0}$ and 
$f(\pm 0)=\pm \Delta m$ but $f(0)=0$. 

The Lorentz invariant non-local factor  in (3) after the modification in (19), 
\begin{eqnarray}\label{(27)}
\left[\theta(x^{0}-y^{0})-\theta(y^{0}-x^{0})\right]
\left[\delta((x-y)^{2}-l^{2})-\delta((x-y)^{2})\right],
\end{eqnarray}
cancels out the infinite time-like volume effect and eliminates the quadratic
infrared divergence completely. In effect, the non-locality is limited to the fluctuation around the tip of the light-cone characterized by the length scale $l$, which we choose to be the Planck length.
 
By setting 
\begin{eqnarray}
l=1/M_{P},\ \ \  \mu=M^{3},
\end{eqnarray}
with the Planck mass $M_{P}$, the particle-antiparticle mass splitting is given by 
\begin{eqnarray}
2\Delta m = 2\pi^{2} \mu l^{2}= 2\pi^{2}M(M/M_{P})^{2}.
\end{eqnarray}
If one chooses $M\sim 10^{9}$ GeV, the particle-antiparticle  mass splitting becomes of the order of the observed neutrino mass (difference) $\sim 0.1$ eV~\cite{particle-data}.

\section{Quantization}
As for the quantization, we adopted the path integral formulation on the basis of Schwinger's action principle in~\cite{CFT1},
 which formally integrates the equations of motion~\cite{fujikawa}.  This scheme which emphasizes the equations of motion  is analogous to the Yang-Feldman formalism~\cite{yang-feldman} in the operator formulation.  We adopt this path integral formulation throughout this paper, since the conventional canonical quantization is not defined for a theory non-local in time.

For the non-locality of the order of the Planck length,  our Lorentz invariant non-local CPT breaking factor (24) is written as
 \begin{eqnarray}\label{(22)}
f(p) = \Delta m[\theta(p_{0})-\theta(-p_{0})]\theta(p^{2}).
\end{eqnarray}
The propagator of the fermion in path integral on the basis of Schwinger's action principle is then given by~\cite{fujikawa}, 
\begin{eqnarray}\label{(31)}
\langle T^{\star}\psi(x)\bar{\psi}(y)\rangle
&=&\int \frac{d^{4}p}{(2\pi)^{4}} e^{-ip(x-y)}\frac{i}{\pslash-m +i\epsilon-\Delta m[\theta(p_{0})-\theta(-p_{0})]\theta(p^{2})},\nonumber\\
\end{eqnarray}
where $T^{\star}$ stands for the covariant T-product which avoids the precise coincident time $x^{0}=y^{0}$.  This propagator shows that the extra  terms are ignored for $p=p_{0}\rightarrow\pm\infty$.  In fact one can confirm that~\cite{FT} 
\begin{eqnarray}
&&f(p)=i[f_{+}(p)-f_{-}(p)]\rightarrow 0, 
\end{eqnarray}
for $p=p_{0}\rightarrow\pm\infty$ in Minkowski space without any approximation.  The propagator for Minkowski momentum is thus well behaved and the effects of non-locality are mild and limited.
If one recalls that the vanishing of
\begin{eqnarray}
\lim_{p_{0}\rightarrow large}\frac{i}{\pslash-m +i\epsilon-\Delta m[\theta(p_{0})-\theta(-p_{0})]\theta(p^{2})}=0
\end{eqnarray}
is the criterion of the T-product in (30) in {\em ordinary theory}, our non-local action almost defines the canonical T-product.  
In the Bjorken-Johnson-Low
prescription, however, one generally requires the stronger condition of vanishing for
$|p_{0}|\rightarrow \infty$ in the complex plane of $p_{0}$~\cite{fujikawa}.

If one considers the Euclidean amplitude obtained from the Minkowski amplitude by Wick rotation, 
our propagator, which contains trigonometric functions, has undesirable behavior after the Wick rotation such as 
\begin{eqnarray}
\sin p_{0}z\rightarrow i\sinh p_{4}z,
\end{eqnarray}
and the exponentially divergent behavior is generally induced in the extra terms. In this sense, the effects of non-locality become non-negligible. 
However, one might still argue that higher order effects in field theory defined in Minkowski space are in principle analyzed in Minkowski space and, if that is the case, our propagator suggests the ordinary renormalizable behavior.

This issue is left for future study.

\section{ Neutrino-antineutrino  mass splitting}

The neutrino masses are not precisely specified by the original
Standard Model and thus may provide a window to a ``brave New World".  It may be allowed to entertain the idea of the possible CPT breaking in the neutrino mass sector and apply our scheme of Lorentz invariant CPT breaking.  Since the Standard Model is very successful, we incorporate the basic properties:\\
a) Lorentz invariance;\\
b) $SU(2)\times U(1)$ gauge invariance;\\
c) C, CP and CPT breaking;\\
d) Non-locality within a distance scale of the Planck length.  
\\

The Standard Model Lagrangian relevant to our discussion of the electron sector is given by
\begin{eqnarray}
{\cal L}&=&i\overline{\psi}_{L}\gamma^{\mu}
\left(\partial_{\mu} - igT^{a}W_{\mu}^{a}
             - i\frac{1}{2}g^{\prime}Y_{L}B_{\mu}\right)\psi_{L}
\nonumber\\
         & +&i\overline{e}_{R}\gamma^{\mu}(\partial_{\mu}
             + ig^{\prime}B_{\mu})e_{R}
+i\overline{\nu}_{R}\gamma^{\mu}\partial_{\mu}\nu_{R}\nonumber
\\
            &+&[ -
\frac{\sqrt{2}m_{e}}{v}\overline{e}_{R}\phi^{\dagger}\psi_{L}
-\frac{\sqrt{2}m_{D}}{v}\overline{\nu}_{R}\phi_{c}^{\dagger}\psi_{L}  -\frac{m_{R}}{2}\nu_{R}^{T}C\nu_{R} + h.c.],
\end{eqnarray}
with the {\em assumed} $\nu_{R}$. We denote the 
Higgs doublet and its $SU(2)$ conjugate by $\phi$  and $\phi_{c}\equiv i\tau_{2}\phi^{\star}$.
We tentatively set $m_{R}=0$ with enhanced lepton number symmetry, namely,  the "Dirac neutrino";  we assume that  every mass arises from the observed Higgs doublet.

One may  add a hermitian  non-local Higgs coupling with a real parameter $\mu$ to the Lagrangian~\cite{FT},
\begin{eqnarray}\label{(5)}
{\cal L}_{CPT}(x)
&&=-i\frac{2\sqrt{2}\mu}{v}\int
d^{4}y\Delta_{l}(x-y)\theta(x^{0}-y^{0})\nonumber\\
&&\times\{\bar{\nu}_{R}(x)\left(\phi_{c}^{\dagger}(y)\psi_{L}(y)\right)
-\left(\bar{\psi}_{L}(y)\phi_{c}(y)\right)\nu_{R}(x)\},
\end{eqnarray}
without spoiling Lorentz invariance and $SU(2)_{L}\times U(1)$ gauge symmetry with $\Delta_{l}(x-y)$ defined in (19).
 
In the unitary gauge, the neutrino mass  term 
becomes
\begin{eqnarray}\label{mass}
S_{\nu \rm mass}
&=&\int d^{4}x\Big\{-m_{D}\bar{\nu}(x)\nu(x)\left(1+\frac{\varphi(x)}{v}\right)\nonumber\\
&& -i\mu\int
d^{4}y\Delta_{l}(x-y)\left[\theta(x^{0}-y^{0})-\theta(y^{0}-x^{0})\right]\bar{\nu}(x)\nu(y)\nonumber\\
&&+i\mu\int
d^{4}y\Delta_{l}(x-y)\bar{\nu}(x)\gamma_{5}\nu(y)
\\
&&-i\frac{\mu}{v}\int
d^{4}y\Delta_{l}(x-y)\theta(x^{0}-y^{0})\left[\bar{\nu}(x)(1-\gamma_{5})\nu(y)-\bar{\nu}(y)(1+\gamma_{5})\nu(x)\right]\varphi(y)\Big\},\nonumber
\end{eqnarray}
and the term 
\begin{eqnarray}\label{(8)}
-i\mu\int
d^{4}x\int d^{4}y\Delta_{l}(x-y)
\left[\theta(x^{0}-y^{0})-\theta(y^{0}-x^{0})\right]
\bar{\nu}(x)\nu(y)
\end{eqnarray}
in the action  preserves $T$ but has $C=CP=CPT=-1$ and thus gives rise to particle-antiparticle mass splitting~\cite{FT}.

For  time-like $p^{2}>0$, one may go to the frame where $\vec{p}=0$ and obtain the eigenvalue equation
\begin{eqnarray}\label{(14)}
p_{0}&=&\gamma^{0}[m_{D}+f(p_{0})+ig(p^{2}_{0})\gamma_{5}],
\end{eqnarray}
with
\begin{eqnarray}\label{(15)}
f(p_{0})
&\equiv&i[f_{+}(p_{0})-f_{-}(p_{0})]\nonumber\\
&=&-4\mu\pi\int_{0}^{\infty}dz\Big\{\frac{z^{2}\sin [p_{0}\sqrt{z^{2}+l^{2}}]}{\sqrt{z^{2}+l^{2}}}
-\frac{z^{2}\sin [p_{0}\sqrt{z^{2}}]}{\sqrt{z^{2}}}\Big\}
\end{eqnarray}
and
\begin{eqnarray}\label{(16)}
g(p^{2}_{0})&=&-4\mu\pi\int_{0}^{\infty}dz\Big\{\frac{z^{2}\cos [p_{0}\sqrt{z^{2}+l^{2}}]}{\sqrt{z^{2}+l^{2}}}-\frac{z^{2}\cos [p_{0}\sqrt{z^{2}}]}{\sqrt{z^{2}}}\Big\}.
\end{eqnarray}
Since we are assuming that the $CPT$ breaking terms are small, we may
solve the mass eigenvalue equations iteratively
\begin{eqnarray}\label{(17)}
m_{\pm}
&\simeq&m_{D}+i\gamma_{5} g(m_{D}^{2}) \pm
f(m_{D}).
\end{eqnarray}
The parity violating mass $+i\gamma_{5} g(m_{D}^{2})$ is now transformed away by a suitable
global chiral transformation. 

In this way, the
neutrino-antineutrino mass splitting is incorporated in the
Standard Model through the Lorentz invariant non-local $CPT$ breaking
mechanism, without spoiling the $SU(2)_{L}\times U(1)$ gauge symmetry. 
The neutrino-antineutrino mass splitting is given by the formula in (28)
\begin{eqnarray}
2\Delta m = 2\pi^{2} \mu l^{2}= 2\pi^{2}M(M/M_{P})^{2}.
\end{eqnarray}
Thus the neutrino-antineutrino mass splitting 
\begin{eqnarray}
\Delta m=10^{-1}\sim 10^{-2} eV,
\end{eqnarray}
which is intended to be  of the order of $m_{D}/5$, is generated by $M\simeq 10^{8}\sim 10^{9}$ GeV and  appears to be allowed by presently available experimental data such as MINOS~\cite{minos}.

As for the induced $CPT$ violating effect on the electron-positron splitting, it  is shown to be {\em finite} (for the one-loop correction) and estimated at the order~\cite{FT}
\begin{eqnarray}\label{(39)}
\alpha [m_{D}m_{e}/M_{W}^{2}](\mu l^{2})[\theta(k^{0})-\theta(-k^{0})] \theta(k^{2}),
\end{eqnarray}
which, for $\Delta m =\pi^{2}\mu l^{2}=10^{-1}\sim 10^{-2}$ eV, is 
\begin{eqnarray}
|m_{e}-m_{\bar{e}}|\sim 10^{-20} eV,
\end{eqnarray} 
and thus well below the present experimental upper bound $\leq 10^{-2}$ eV~\cite{particle-data}.  We can thus avoid the rather involved issue of gauge invariance which needs to be analyzed  for the case of electron-positron mass splitting~\cite{CFT2}; this conclusion comes with the qualification that the calculation is based on a Minkowski space evaluation using (30) and thus it requires a future re-examination, although the effect is in any case expected to be very small.

The induced $CPT$ violation is expected to be smaller in the quark sector (as a two-loop effect) than in the charged leptons in the $SU(2)\times U(1)$ invariant theory, and thus much smaller than the well-known limit on the $K$-meson~\cite{particle-data}, 
\begin{eqnarray}
|m_{K}-m_{\bar{K}}|<0.44\times 10^{-18} GeV.
\end{eqnarray}

\section{Conclusion}

We illustrated a possible CPT violation in a  Lorentz invariant non-local theory which gives rise to a mass splitting between the particle and antiparticle. The full quantum mechanical treatment of this specific non-local theory has not been analyzed, but the lowest order one-loop corrections are finite and thus promising.  Ultimately, the consistency of the present scheme is expected to be related to our understanding of space-time at the Planck length. This model is applied to the possible neutrino-antineutrino mass splitting in the Standard Model.

The test of CPT symmetry in neutrino oscillation has been discussed in some detail in~\cite{pakvasa}.  More concrete analyses of a
possible sizable  neutrino-antineutrino mass splitting have been performed  in connection with the LSND experiment in~\cite{murayama}, but such a large mass splitting of the order of $1 {\rm eV}$ appears to be disfavored by experiments~\cite{minos}.  An implication of the possible neutrino mass splitting on the baryon asymmetry was also discussed~\cite{smirnov}\cite{ FT}.  At this moment, it is known that there is a small mass discrepancy  between the antineutrino in the reactor experiment and the neutrino from the Sun, which is about $2\sigma$ discrepancy~\footnote{ Private communications from A. Suzuki and A.Y. Smirnov.}.  This may indicate that the neutrino oscillation is a good testing ground of CPT symmetry.


\begin{thebibliography}{99}
\bibitem{weinberg}
S. Weinberg, Phys. Rev. Lett. {\bf 19} (1967) 1264.
\bibitem{salam}
A. Salam, in: N. Svartholm (ed.), {\em Elementary Particle Theory}, Stockholm, 1968, p.367.
\bibitem{glashow}
S.L.  Glashow, Nucl. Phys. {\bf 22} (1961) 579. 
\bibitem{pauli}
W. Pauli, {\it Niels Bohr and the Development of Physics}, W. Pauli
(ed.),  Pergamon Press, New York, 1955.
\bibitem{luders}
G. L\"{u}ders, 
Mat. Fys. Medd. Dan. Vid. Selsk.	{\bf 28}, no. 5 (1954) 1.
\bibitem{piguet}
O.M. Del Cima, D.H.T. Franco, A.H. Gomes, J.M. Fonseca, O. Piguet, Phys. Rev. D{\bf 85} (2012) 065023, and references quoted therein. 
\bibitem{FT}
K. Fujikawa and A. Tureanu, Phys. Lett. B{\bf 743} (2015) 39; Mod. Phys. Lett. A{\bf 30}  (2015) 1530016.
\bibitem{CFT1}
M. Chaichian, K. Fujikawa and A. Tureanu,
 Phys. Lett. B{\bf 712} (2012) 115. 
\bibitem{CFT2}
M. Chaichian, K. Fujikawa and A. Tureanu, 
 Phys. Lett. B{\bf 718} (2012) 178. 
;\ Phys. Lett. B{\bf 718} (2013) 1500. 
\bibitem{particle-data}
K.A. Olive et al. (Particle Data Group), Chin. Phys. C{\bf  38} (2014)  090001, and 2015 update. 
\bibitem{fujikawa}
K. Fujikawa, Phys. Rev. D {\bf 70} (2004) 085006. 
As for the
Bjorken-Johnson-Low method, see also Appendix in
K. Fujikawa and P. van Nieuwenhuizen, Annals Phys. {\bf 308} (2003) 78. 
\bibitem{yang-feldman}
C.N. Yang and  D. Feldman, Phys. Rev. {\bf 79}  (1950) 972.
\bibitem{pakvasa}
V.D. Barger, S. Pakvasa, T.J. Weiler and K. Whisnant, Phys. Rev. Lett. {\bf 85} (2000) 5055. 
\bibitem{murayama}
H. Murayama and T. Yanagida,  Phys. Lett. B {\bf 520} (2001) 263. 
\\
G. Barenboim, L. Borissov and J. Lykken, Phys. Lett. B {\bf534} (2002) 106. 
\\
S. M. Bilenky, M. Freund, M. Lindner, T. Ohlsson and W. Winter,  Phys.
Rev. D {\bf65} (2002) 073024. 
\\
G.~Barenboim and J.~D.~Lykken,
   Phys.\ Rev.\ D {\bf 80} (2009) 113008. 
\bibitem{minos}
P.~Adamson {\it et al.}  [MINOS Collaboration],
Phys. Rev. Lett. {\bf 108} (2012) 191801.
\bibitem{smirnov}
G. Barenboim, L. Borissov, J. D. Lykken and A. Y. Smirnov, JHEP {\bf 0210} (2002)  001. 


\end{thebibliography}
\end{document}